\documentclass[aip,reprint,jcp]{revtex4-1}
\usepackage{amsmath}
\usepackage{amssymb}
\usepackage[dvips]{graphicx}
\usepackage[svgnames]{xcolor}
\usepackage[colorlinks=true,citecolor=blue,linkcolor=blue,citebordercolor=white,linkbordercolor=white]{hyperref}
\usepackage{natbib}

\def\eq#1{{Eq. (\ref{#1})}}

\begin{document}

\title[Charge transport in doped polycarbonate]{Charge carrier transport in molecularly doped polycarbonate as a test case for the dipolar glass model}

\author{S.V. Novikov}
\email{novikov@elchem.ac.ru}
\affiliation{A.N. Frumkin Institute of
Physical Chemistry and Electrochemistry, Leninsky prosp. 31, Moscow 119991, Russia}
\affiliation{National Research Nuclear University MEPhI, Kashirskoe Sch. 31, Moscow 115409, Russia}
\author{A.P. Tyutnev}
\affiliation{National Research University Higher School of Economics, Myasnitskaya Ulitsa 20, Moscow 101000, Russia}

%\pacs{73.50.Pz,73.61.Ph}

\begin{abstract}
We present the results of Monte-Carlo simulations of the charge carrier transport in a disordered molecular system containing spatial and energetic disorders using the dipolar glass model. Model parameters of the material were chosen to fit a typical polar organic photoconductor polycarbonate doped with 30{\%} of aromatic hydrazone, whose transport properties are well documented in literature. Simulated carrier mobility demonstrates a usual Poole-Frenkel field dependence and its slope is very close to the experimental value without using any adjustable parameter. At room temperature transients are universal with respect to the electric field and transport layer thickness. At the same time, carrier mobility does not depend on the layer thickness and transients develop a well-defined plateau where the current does not depend on time, thus demonstrating a non-dispersive transport regime. Tails of the transients decay as power law with the exponent close to -2. This particular feature indicates that transients are close  to the boundary between dispersive and non-dispersive transport regimes. Shapes of the simulated transients are in very good agreement with the experimental ones. In summary, we provide a first verification of a self-consistency of the dipolar glass transport model, where major transport parameters, extracted from the experimental transport data, are then used in the transport simulation, and the resulting mobility field dependence and transients are in very good agreement with the initial experimental data.
\end{abstract}
\maketitle

\section{Introduction}

Dipolar glass (DG) model provides a natural and attractive description of major transport properties of polar amorphous organic materials. Initially, it was suggested for an explanation of the universal experimental observation of the Poole-Frenkel (PF) mobility field dependence
\begin{equation}\label{mu_PF}
\ln\mu=\textrm{const}+bE^{1/2}
\end{equation}
in amorphous organic materials in a broad field range spanning almost two decades $10^4-10^6$ V/cm. \citep{Borsenberger:9,Borsenberger:book,Pope:1328} This model could be considered as a further development of the Gaussian Disorder Model (GDM), \citep{Bassler:15} which assumes that charge transport in organic materials occures as a series of hops in random Gaussian energy landscape $U(\vec{r})$, created by contributions from randomly packed molecules. Long ago Borsenberger and B{\"a}ssler  \citep{Borsenberger:5327} demonstrated  that the GDM explains a strong polarity effect on carrier mobility by taking into account the dipolar as well as van der Waals energetic disorder. Yet, by the very nature of the GDM, it does not incorporate the most striking feature of the dipolar disorder, i.e. a long range spatial correlation of the random energy landscape,\cite{Novikov:14573,Dunlap:542} and for this reason cannot adequately explains emergence of the PF dependence. In contrast with the GDM, DG model explicitly takes into account spatial correlations and naturally provides the PF field dependence in a wide field range. \cite{Dunlap:542,Novikov:130,Novikov:4472,Dunlap:437,Novikov:2584}

Initially, the DG model has been dubbed "Correlated Disorder Model" (CDM)\cite{Novikov:4472}, stressing the most important qualitative difference of the model with the GDM, but the proposed name describes better the true nature of the model. Indeed, as it was shown in Ref. \onlinecite{Dunlap:542}, different spatial correlations of the random energy landscape lead to the different field dependences of the mobility, hence the CDM name is too vague. The particular model, suggested and studied in Refs. \onlinecite{Dunlap:542,Novikov:4472}, is best suited for the description of transport properties of polar organic materials containing molecules having permanent dipole moments.

For the study of hopping charge transport in organic glasses a lattice version of the DG model is usually considered with sites of the regular lattice occupied by randomly oriented dipoles, and some of sites are considered as transport ones. Earlier transport simulations for the DG model have been done for the case of the totally filled lattice where the fraction $c$ of sites occupied by transport molecules was equal to 1. Such simulations suggested that the temperature and field dependence of the mobility may be described by the phenomenological equation
\begin{eqnarray}\label{simul}
\ln\mu/\mu_0=-\left(\frac{3\hat{\sigma}}{5}\right)^2+
C_0\left(\hat{\sigma}^{3/2}-\Gamma\right)\left(eaE/\sigma\right)^{1/2},\\
\nonumber \hat{\sigma}=\sigma/kT,
\end{eqnarray}
with $C_0\approx 0.78$ and $\Gamma\approx 2$.\cite{Novikov:4472} Here $\sigma$ is the magnitude of the energetic dipolar disorder and $a$ is the lattice scale. This particular relation has been widely used for a description of transport properties of various materials, we mention here just some recent references.\cite{Kreouzis:235201,Gambino:157,Zufle:63306,Barard:13701,Gadisa:3126,Fong:094502,Vijila:8} In some papers a comparative analysis of the data using both the GDM and DG model was carried out and it was found that the DG model provides more coherent description of charge transport.\cite{Kreouzis:235201,Gadisa:3126,Fong:094502}
Still, a proper experimental test of the DG model is far from completed.

First, at the moment no simulation has been carried out for $c<1$. The lattice model with $c=1$ is adequate for description of low molecular weight organic glasses, while polymeric materials, especially molecularly doped polymers, are certainly better described by the case $c<1$. In this paper we provide results of the transport simulation for a partially filled lattice with $c=0.3$ to verify the validity of the PF relation for that case. Numerous experiments demonstrate that the PF relation is generally valid for $c\simeq 0.2- 1$; in addition, there is no theoretical reason to doubt its validity. Indeed, a general picture of the carrier hopping in the correlated energy landscape suggests that  major features of quasi-equilibrium non-dispersive transport is  governed by the critical valleys with the size\cite{Dunlap:437}
\begin{equation}\label{Rc}
r_c\simeq \left(\frac{\sigma^2 a}{ekTE}\right)^{1/2}.
\end{equation}
 If the typical distance between transport sites $r_t\simeq a/c^{1/3}$ is much less than $r_c$
\begin{equation}\label{condition}
r_c \gg r_t \hskip10pt \textrm{or}\hskip10pt c\gg \left(\frac{eakTE}{\sigma^2}\right)^{3/2},
\end{equation}
one should not expect any significant modification of the PF dependence. It is worth noting that according to this line of reasoning, with the decrease of $c$ the mobility dependence initially begins to deviate from the strict PF law in the high field region. For $E\approx 1\times 10^6$ V/cm, $\sigma\simeq 0.1$ eV, and $a\simeq 1$ nm inequality (\ref{condition}) gives $c \gg 0.1$ at room temperature. Nonetheless, a direct check of the validity of relation (\ref{mu_PF}) for a partially filled lattice is highly desirable.

Next, the most important reason to perform  the simulation for $c<1$ that uses  set of transport parameters extracted from experimental data for a particular transport material is to provide a test of the self-consistency of the transport model. For quite a long time the GDM and DG model are used for the description of transport properties of organic materials, i.e. relevant transport parameters (mostly $\sigma$) are calculated using experimental mobility data according to the prescriptions of the particular model. All such prescriptions are provided by the phenomenological relations, based on the extensive Monte Carlo simulations.\cite{Bassler:15,Novikov:4472} In this situation a very natural suggestion should be to perform a simple but very convincing test for the self-consistency of the model: to carry out the simulation for a given set of parameters (experimental ones) and then compare the results (mostly, the simulated mobility field dependence) with the experimental data. To the best of our knowledge, such comparison has been done neither for the GDM nor DG model. In this paper we present the result of such test for the particular set of transport parameters, extracted from the experimental mobility data for a typical organic semiconducting glass containing 30\% (by weight) hole transporting aromatic hydrazone DEH in polycarbonate (PC) matrix, and then compared our results with the experimental ones.\cite{Schein:686,Schein:41,Schein:1067,Dunlap:9076,Tyutnev:115107} Experimental results, provided in different papers, agree well, thus justifying our choice of the reference transport material.

Last, earlier Monte-Carlo simulations \citep{Novikov:444} demonstrated that time-of-flight (TOF) current transients, predicted by the DG model, reproduce very well the experimental transients: there is a short spike reflecting initial spatial and energetic relaxation of carriers, followed by a flat plateau signaling emergence of the quasi-equilibrium  transport with the mean carrier velocity independent of time, and then a final tail marking the arrival of carriers to the collecting electrode and typically having algebraic form $I(t)\propto 1/t^{\beta}$ with $\beta\simeq 2.0-2.5$.

At the same time, a general behavior of transients was not well studied because the paper\citep{Novikov:444} was entirely devoted to the study of an important but very particular effect of deep traps on the shape of transients.  In this paper we study TOF current shapes predicted by the DG model via Monte Carlo simulation for a partially filled lattice and compare simulated transients with the experimental ones for 30\% DEH:PC. Again, we use this comparison as a test of the self-consistency of the DG model.

Realization of this program could provide the DG model with more strong footing.

\section{Basics of the DG model and simulation details}

Major assumptions of the DG model are well known. \citep{Novikov:14573,Dunlap:542,Dunlap:437,Novikov:2584} Usually the model is considered as a regular cubic lattice with sites occupied by randomly oriented dipoles having dipole moment $p$. We assume that orientation of dipoles is static and not affected by the applied uniform electric field $E$. The dipoles generate a random spatial distribution of the electrostatic potential  $\varphi(\vec{r})$ which immediately translates to the random site energy $U(\vec{r}) = e\varphi (\vec{r})$ with zero mean and rms $\sigma= \left<U^2(\vec{r})\right>^{1/2}= 0.05 - 0.1$ eV, depending on $a$ and $p$.\citep{Novikov:14573,Dunlap:542} Particular energy distribution was obtained by the summation of contributions of all dipoles using the Ewald method\cite{DeLeeuw:27} and assuming the periodic boundary conditions for the distribution of dipoles in the basic sample; we assumed also that every site is occupied by a dipole. If the fraction of sites, occupied by dipoles, is not very low and the average distance between neighbor dipoles is comparable to $a$, then the resulting distribution of $U$ has a Gaussian form; for low concentration of dipoles the tail of distribution decays more slowly. \cite{Dieckmann:8136,Novikov:877e,Novikov:14573} If the energy landscape has a Gaussian distribution, then the only relevant parameter is $\sigma$; fraction of sites, occupied by dipoles, the number of different sorts of dipoles, and the very nature of dipoles (i.e., are they transport sites or not) are irrelevant for the problem.

Random energy landscape provided by dipoles has a particular property which is in drastic contrast to the GDM: a binary correlation function $C(\vec{r}) = \left<U(\vec{r})U(0)\right>$ decays slowly with distance,  $C(\vec{r})\propto   a/r$, $r \gg a$, \citep{Novikov:14573,Dunlap:542} while for the GDM this function is zero everywhere apart from $\vec{r} = 0$, where  $C(0) =  \sigma^2$. Long range correlation naturally leads to the PF dependence: the 1D transport model predicts that if $C(\vec{r})\propto 1/r^n$, then $\ln\mu\propto E^{n/(n+1)}$. \cite{Dunlap:542}

In typical organic polar materials the actual random energy landscape $U(\vec{r})$ incorporates contributions from other sources of energetic disorder: quadrupolar,  \cite{Novikov:89,Novikov:954} van-der-Waals, \cite{Borsenberger:11314} etc. The dipolar contribution stands apart because its correlation function  decays more slowly than corresponding functions for other sources of disorder.\cite{Dunlap:80,Novikov:2584} This means that the dipolar disorder, when not negligible, provides the dominant contribution to the mobility field dependence for low and moderate fields.\cite{Dunlap:542,Novikov:2532} This fact justifies the use of a pure dipolar disorder (i.e., the DG model) for description of the mobility field dependence in polar amorphous organic materials for a field range $10^4-10^6$ V/cm.

Our simulation technique is very close to the one, described previously.\cite{Novikov:4472} The only difference is that the fraction of sites, occupied by transport molecules is not $c=1$ but $c=0.3$, thus providing connection with the experimental data for a particular well-characterized  system, 30\% DEH:PC.  \cite{Schein:686,Schein:41,Schein:1067,Dunlap:9076,Mack:7500} We used the simplest assumption that sites of the lattice are occupied by transport molecules independently with  probability $c$; the use of the simplest hypothesis is mostly related to the lack of relevant experimental information on the local structure of organic glasses.

In our simulations we examine a temporal dependence of the carrier velocity $v(t)$, averaged over many carriers for a particular realization of the random energy landscape $U(\vec{r})$, and then over realizations of  $U(\vec{r})$.   Current transient $I(t)$ in the TOF experiment is directly proportional to $v(t)$. Another important transport parameter, used for the calculation of the Monte Carlo carrier mobility, is a mean carrier velocity $\left<v\right>=\left<L/t_\textrm{drift}\right>$, where $t_\textrm{drift}$ means the time for a carrier to reach the collecting electrode.

We used the Miller-Abrahams (MA) hopping rate,\cite{Miller:745} where the rate of transition from site $i$ to site $j$ is given by
\begin{equation}\label{MA_rate}
p_{i\rightarrow j}=\nu_0 \exp(-2\gamma r_{ij})\begin{cases}
    \exp\left(-\frac{U_j-U_i}{kT}\right), &U_j-U_i > 0\\1, &U_j-U_i < 0
\end{cases}
\end{equation}
where $\nu_0$ is the prefactor frequency, $r_{ij}=|\vec{r}_j-\vec{r}_i|$, and $\gamma$ is a wave function decay parameter for transport sites; in organic materials $\gamma a\simeq 5-10$.\cite{Bassler:15,Pasveer:206601} In the presence of applied electric field $E$ the site energy $U_i$ includes an additional term $-e\vec{r}_i\vec{E}$. It was found previously that details of the hopping rate are irrelevant to the emergence of the PF dependence (\ref{mu_PF}), at least for the case $c=1$. \cite{Novikov:4472}

Relevant physical parameters for the simulation were taken from the experimental transport data for 30 wt\% DEH:PC glass at room temperature.\cite{Schein:686,Schein:41,Schein:1067} We used $\sigma=0.13$ eV, $a=7.7$\AA, $2\gamma a=11.8$, and $kT=0.0252$ eV. In these papers $\sigma$ was calculated using the GDM analysis of the low field mobility. The corresponding DG value $\sigma_\textrm{DG}$ could be obtained multiplying by the correction factor 10/9. \cite{Bassler:15,Novikov:4472} We did not perform this correction because it is well known that the coefficient $A$ in the relation
\begin{equation}\label{low_E}
   \ln(\mu/\mu_0)\left.\right|_{E\rightarrow 0} = -A \left(\sigma/kT\right)^2,
\end{equation}
valid for both GDM and DG model, is not a constant but depends on $\gamma a$.\cite{Parris:218} This dependence is rather weak around $\gamma a\approx 5-10$, but, nonetheless, it provides some uncertainty for a value of $\sigma$, calculated from the experimental mobility data. In addition, there may be a weak dependence of $A$ on $c$ as well. In this situation it is difficult to make a reliable but rather small correction from the GDM to DG model.

Exceptions are Figs. \ref{PF_exp} and \ref{universal144}, where simulation was performed using $\sigma_\textrm{corr}=10/9\thinspace\sigma=0.144$ eV as well, in order to estimate the agreement between our simulation and experimental mobility data for 30\% DEH:PC. Simulation time becomes prohibitively long to obtain good quality transients (especially long-time tails) using  $\sigma_\textrm{corr}$, while the corresponding simulation of a more robust integral transport parameter (i.e., the mobility) still remains feasible. Nonetheless, there is no doubt that the major features of transients (such as universality, see Section \ref{univ}) remain intact for the case of corrected $\sigma$.

%==================================================
Quite frequently a polaron (or Marcus) hopping rate is considered as a more realistic alternative to the MA   rate\cite{Parris:126601,Baldo:085201}
\begin{equation}\label{polaron_rate}
p_{i\rightarrow j}\propto \frac{1}{(E_b kT)^{1/2}}\exp\left[-\frac{\left(U_j-U_i+2E_b\right)^2}{8E_b kT}\right],
\end{equation}
where $E_b$ is the polaron binding energy and we omit here the dependence on  distance. For typical organic molecules used in transport materials intramolecular contribution to $E_b$ is 100-150 meV,\cite{Coropceanu:926}
and comparable intermolecular contribution could be expected. In this situation a major difference with the MA rate is an additional contribution $E_b/2$ to the effective activation energy $2A\sigma^2/kT$. Significant effect on the shape of transients (as well as on the mobility field dependence) could be expected for rather exotic case of small $E_b\simeq 10-30$ meV, where transport is dominated by the quadratic term $\left(U_j-U_i\right)^2$ in \eq{polaron_rate} and occurs in the inverted regime.\cite{Parris:126601} This statement is supported by our preliminary data. Detailed study of transients for the polaron hopping rate will be published elsewhere. At the same time, a possible polaronic contribution to the transport activation energy provides an uncertainty for the value of $\sigma$, estimated from the experimental data. Unfortunately, at the moment there is no reliable method to extract polaronic contribution to the mobility temperature dependence. For this reason we assumed that all activation dependence originates from the random energy landscape.

%==============================================================

In order to check the sensitivity of the simulation data to a finite size effect, mentioned in Refs. \onlinecite{Lukyanov:193202,Kim:1897}, we performed simulation for basic samples with size $N^3$ equal to $50^3$, $100^3$, and $150^3$ lattice sites and found that the parameter, most sensitive to $N$, is the absolute value of the mobility. Mobility field dependence and relative shape of the transients are much less sensitive. For example, slope of the mobility field dependence $S=\frac{d\ln\mu}{dE^{1/2}}$ in Fig. \ref{PF_exp} varies by 5\% only and general features of transients remain the same. In addition, the finite size effect has been demonstrated for $E=0$ as a deviation of the mean carrier energy $\varepsilon_N$ from the value $\varepsilon_\infty=-\sigma^2/kT$, which is a  mean carrier energy for $E=0$ and infinite number of sites in the case of the Gaussian distribution; for a finite $N$ $\varepsilon_N > \varepsilon_\infty$. The reason for the deviation is scarcity of sites with low energy. However, for $E > 0$ mean carrier energy increases with $E$ and becomes more close to the maximum of the density of states.\cite{Pautmeier:587} For this reason the finite size effect becomes less severe for the simulation with $E>0$.

We should note two earlier papers where the Monte Carlo simulation of the hopping transport in amorphous organic materials for $c < 1$ has been carried out assuming random  distribution of hopping sites.\cite{Hartenstein:321,Sin:901} Contrary to our approach, simulation in Ref. \onlinecite{Hartenstein:321} has been carried out for the GDM. Sin and Soos \cite{Sin:901} did not simulate transients and considered only the case of moderate disorder $\sigma/kT \le 3$, while for 30\% DEH:PC and room temperature $\sigma/kT > 5$. In addition, they generated the correlated energy landscape in somewhat artificial way according to the method suggested by Gartstein and Conwell.\cite{Gartstein:351} This method gives short range spatial correlation and cannot reproduce the proper spatial correlation in polar organic materials.

Some earlier papers considered the concentration dependence of the hopping mobility in organic materials using various analytic approaches.\cite{Arkhipov:505,Rubel:014206,Baranovskii:1644} Unfortunately, such approaches do not provide any information on the shape of transients.

\section{Results and discussion}

\subsection{Mobility field dependence}
\label{Mfd}

\begin{figure}[tbh]
     \begin{center}
\includegraphics[width=3in]{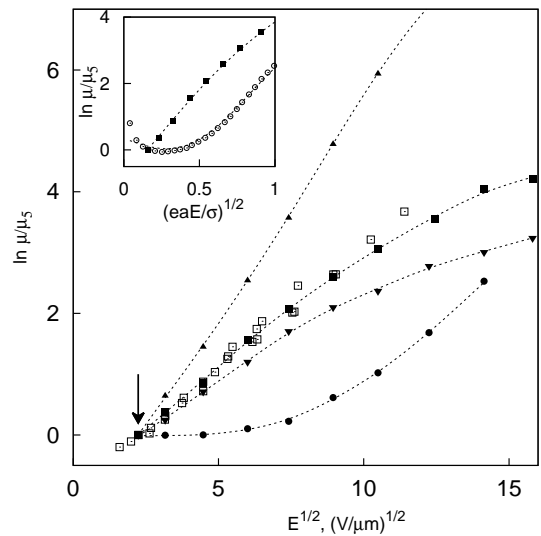}
     \end{center}
\caption{Comparison of the experimental mobility field dependence (denoted by $\Box$, data are borrowed from Refs. \onlinecite{Schein:1067,Tyutnev:115107,Mack:7500}; we do not discriminate points from different papers, they are agree well with each other) and simulated ones (DG model for $\sigma=0.13$ eV ($\blacktriangledown$) and corrected $\sigma_\textrm{corr}=0.144$ eV ($\blacksquare$), and for the GDM ($\bullet$), correspondingly). The topmost curve ($\blacktriangle$) shows the simulation data for the equivalent 100\% filled lattice with lattice parameter equals to $a/0.3^{1/3}$. Plotted is the logarithm of the ratio $\mu/\mu_5$, where $\mu_5$ is the corresponding mobility for $E=5$ V/$\mu$m (matching point is indicated by the arrow). Hence, the absolute value of the mobility is irrelevant, and only the slope of the mobility field dependence is compared. Note that in this figure there are no fitting parameters. Dotted lines are used as guides for an eye. Inset shows our mobility curve for $\sigma_\textrm{corr}=0.144$ eV (hence, $\sigma/kT=5.7$) and the mobility curve for $\sigma/kT=5.8$ ($\circ$), borrowed from Fig. 3 of Ref. \onlinecite{Tonezer:214101}; for $E=5$ V/$\mu$m in our case $(eaE/\sigma)^{1/2}=0.163$ and for the reference mobility value for data from Ref. \onlinecite{Tonezer:214101} we used, instead of $\mu_5$, the mobility, simulated for the same value of $(eaE/\sigma)^{1/2}$.\label{PF_exp}}
\end{figure}

We found that the PF field dependence is valid for $c=0.3$ as well as for a fully filled lattice (Fig. \ref{PF_exp}). Deviation from the PF dependence begins around $1\times 10^6$ V/cm, reflecting the specific feature of the Miller-Abrahams hopping rate \cite{Novikov:4472} and  effect of the decrease of the concentration of transport sites, mentioned in the Introduction. This figure  also shows that the DG model successfully passes the self-consistency test and simulation with parameters, extracted from experimental data, very well reproduce initial experimental field dependence. Simulation using $\sigma_\textrm{corr}$ does not differ significantly from the simulation using $\sigma$, but corresponding curve $\mu(E,\sigma_\textrm{corr})$ even better agrees with the experimental one.

%===== fig1

We also carried out simulation for the equivalent totally filled lattice, i.e. the lattice with $c_e=1$ and $a_e=a/c^{1/3}\approx 1.5 a$ which simulates the same material 30\% DEH:PC in the traditional lattice gas model.\cite{Bassler:15} If we compare average velocities, then for the  partially filled lattice we have $v=\hat{v}a/\tau$, while for the equivalent lattice (for the same physical thickness of the transport layer and electric field $E$) $v_e=\hat{v}_e a_e/\tau_e$, where raw output of the Monte Carlo simulation  $\hat{v}$ is the dimensionless velocity, measured in the units of $a/\tau$ (here $\tau=\nu_0^{-1}\exp\left(2\gamma a\right)$ is the characteristic hopping timescale). For the equivalent lattice $\tau_e=\nu_0^{-1}\exp\left(2\gamma a_e\right)$, and the ratio of merit is
\begin{equation}\label{eqv_v}
   R=\frac{v}{v_e}=\frac{\hat{v}}{\hat{v}_e}c^{1/3}\exp\left[2\gamma a(c^{-1/3}-1)\right].
\end{equation}
If the approximation of the equivalent lattice is valid, then $R\simeq 1$. In our case, for $E=10$ V/$\mu$m we obtain $R=1.85$; this is a reasonably good agreement (note, that $\hat{v}_e/\hat{v}\approx 230$). However, for higher fields the agreement becomes consistently poorer (see the topmost curve in Fig. \ref{PF_exp}). At the same time, almost all discussions of the comparison of transport properties of organic materials with different concentrations of transport sites extensively use, albeit sometimes implicitly, the equivalent lattice approximation (see, for example, recent Refs. \onlinecite{Schein:7295,Novikov:2532}). Conclusions of such discussions should be reexamined.

It is natural to assume that the mobility curve for the totally filled lattice could be made more close to the experimental one by the addition of the effective spatial disorder according to the prescription of Ref. \onlinecite{Bassler:15} because the addition of such disorder leads to the decrease of the slope $S$ of the mobility field dependence. However, in this approach the spatial disorder is  introduced in a pure phenomenological manner without any possibility to predict a magnitude of disorder, necessary for the reproduction of the proper mobility field dependence. Quite the contrary, in our approach a simple use of the proper concentration of transport molecules immediately provides a good description of the mobility field dependence.

Our results also provide good illustration on the importance of long range spatial correlations for the development of the proper mobility field dependence. Sometimes in the literature one can find a statement that short range correlations are quite sufficient for the development of the field dependence which is almost indistinguishable from the true PF dependence in a wide field range and, hence, for a description of transport properties of organic materials (a recent example is provided in Ref. \onlinecite{Tonezer:214101}) In order to understand the importance of long range correlations it is useful to compare our Fig. \ref{PF_exp} (inset) and Fig. 3 of the cited paper. In our case the PF dependence starts immediately from the lowest tested field, while for the short range disorder the slope for the $eaE/\sigma < 0.3$ is almost zero, or even negative, though, probably, the reason for this unusual behavior is small thickness of the simulated transport layer. For correlations that are negligible for distances greater than $3a$ (Fig. 2 of Ref. \onlinecite{Tonezer:214101}) the resulting mobility field dependence has to be very close to the GDM dependence, and this is clearly seen in Fig. \ref{PF_exp}.

\subsection{Current transients}
\label{univ}

\begin{figure}[tbh]
     \begin{center}
\includegraphics[width=3in]{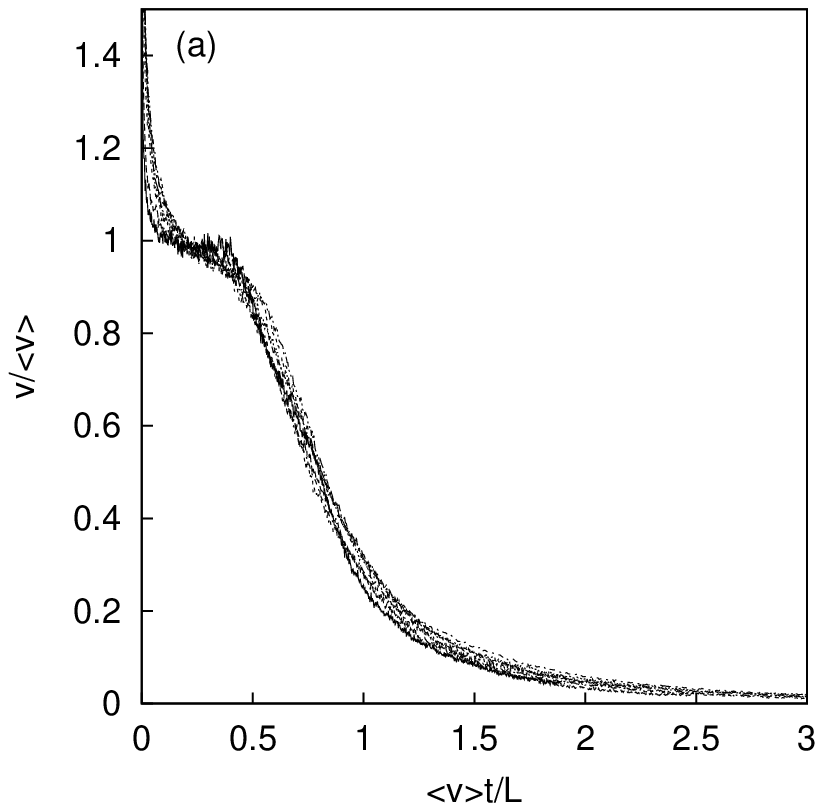}
\includegraphics[width=3in]{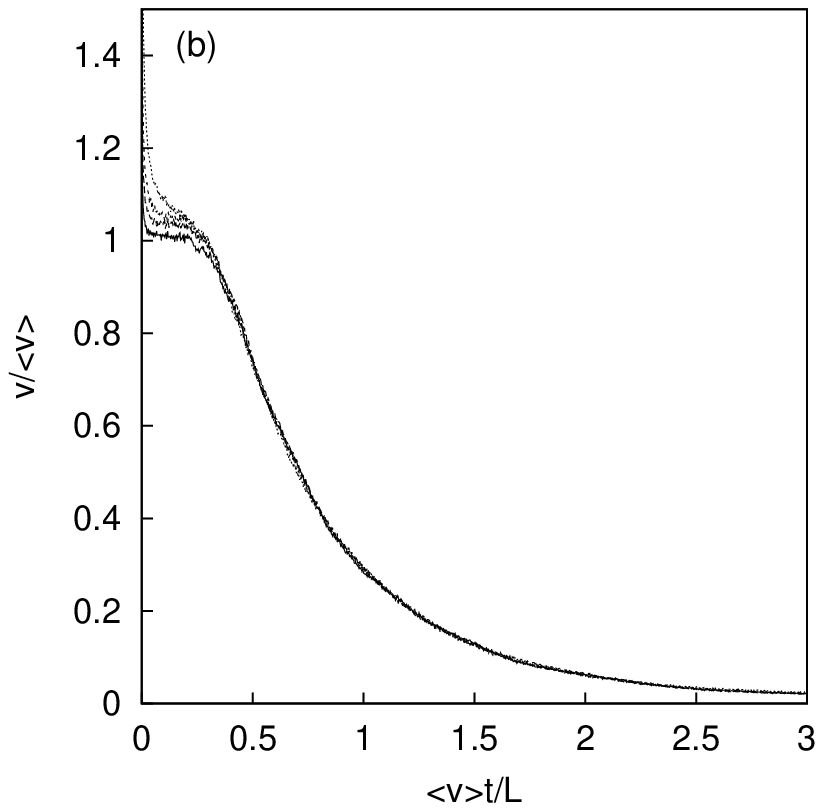}
     \end{center}
\caption{Universality of $I(t)$ (or temporal dependence of $v$): for electric field in the range 5 - 200 V/$\mu$m and $L=20\thinspace 000\thinspace a$, here different lines correspond to the different values of $E$   (a); and for layer thickness in the range $2\thinspace 000\thinspace a$ -- $20\thinspace000\thinspace a$,  $E=5$ V/$\mu$m, here different lines correspond to the different values of $L$ (b). For both plots $kT/\sigma=0.19$ (room temperature). \label{universal}}
\end{figure}
%===== fig2

\begin{figure}[tbh]
     \begin{center}
\includegraphics[width=3in]{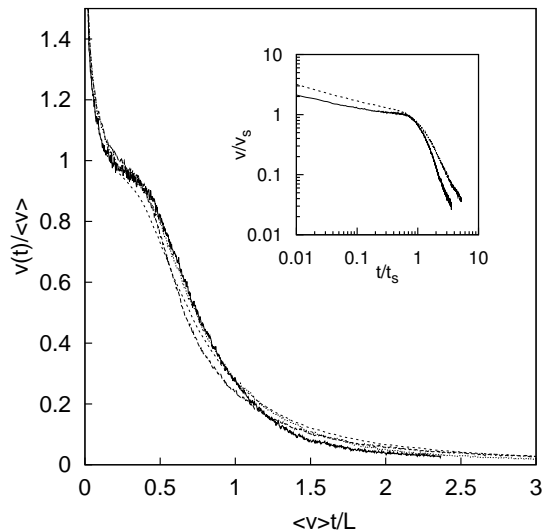}
     \end{center}
\caption{Universality of $v(t)$ for electric field in the range 55 - 200 V/$\mu$m and $L=20\thinspace 000\thinspace a$, here different lines correspond to the different values of $E$; transient are simulated for $\sigma_{\rm corr}=0.144$ eV. Inset shows transients for $E=55$ V/$\mu$m, simulated for the GDM (broken line, $\sigma=0.13$ eV) and DG model (solid line, $\sigma_{\rm corr}=0.144$ eV); $t_s$ and $v_s$ provide a suitable normalization. \label{universal144}}
\end{figure}
%===== fig3

\begin{figure}[tbh]
     \begin{center}
\includegraphics[width=3in]{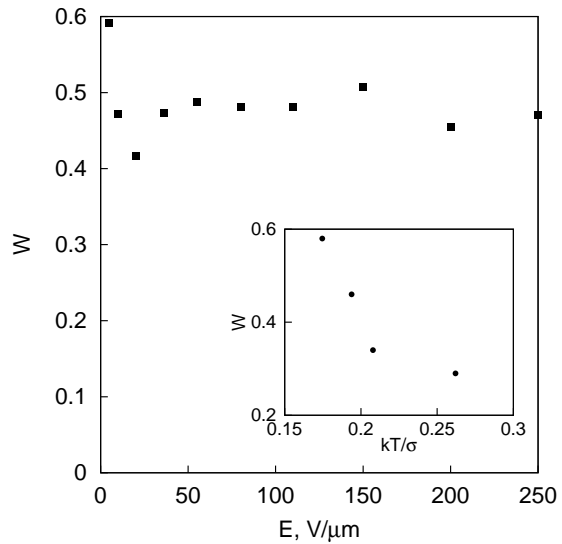}
     \end{center}
\caption{Dependence of the parameter $W$ on $E$ for room temperature; inset shows the dependence of $W$ on $T$ for $E=10$ V/$\mu$m.\label{universal2}}
\end{figure}
%===== fig4

It was already established previously that in our case charge transport occurs in the non-dispersive regime; for this reason mobilities calculated in different ways (as a true Monte Carlo mobility $\mu=\left<v\right>/E$ calculated by mean carrier velocity $\left<v\right>$, or using  two experimental methods of the mobility calculation, estimating either the time of intersection of asymptotes to the plateau and tail of the transient, or the time of the transient to reach one half of the plateau value), are very close.\cite{Novikov:68} Taking into account all relevant parameters for  a non-dispersive regime with $t_\textrm{drift}\propto L$ (here $L$ is a thickness of the transport layer), a general shape of the transient may be written as
\begin{equation}\label{3ps}
    v(t)=\left<v\right>F_3\left(\left<v\right>t/L,kT/\sigma, E/E_s\right),
\end{equation}
i.e., as a three-parameters scaling function, where $E_s$ is some characteristic field. We found that at room temperature ($kT/\sigma\approx 0.19$) transients demonstrate a universality with respect to $L$ and $E$ (see Figs. \ref{universal} - \ref{universal2}), found previously in experiments.\cite{Schein:175,Borsenberger:967,Kreouzis:235201} In agreement with experiment, transients are not universal with respect to $T$. Universality was also confirmed in a more limited field range for the transients simulated using $\sigma_{\rm corr}=0.144$ eV (Fig. \ref{universal144}). In this case transients are more dispersive, yet still less dispersive in comparison to the corresponding GDM transients, calculated for $\sigma=0.13$ eV. Fig. \ref{universal2} shows the field dependence of the parameter
\begin{equation}\label{W}
    W=\frac{t_{1/2}-t_0}{t_{1/2}},
\end{equation}
which provides a simplest robust integral characteristic of the transient shape. Here $t_0$ is the time of the intersection of the asymptotes to the plateau and tail of the transient, and $t_{1/2}$ is the time for the current to decay to one half of the plateau value.
If $E_s$ does not depend on $L$ (i.e., if it is some microscopic field), then \eq{3ps} implies the universality with respect to $L$, in agreement with the simulation result, but the universality with respect to $E$ means in addition that the third parameter in \eq{3ps} is irrelevant, and the two-parametric scaling takes place
\begin{equation}\label{2ps}
    v(t)=\left<v\right> F_2\left(\left<v\right>t/L,kT/\sigma\right),
\end{equation}
where the dependence of $v(t)$ on $E$ is exclusively provided through $\left<v\right>$.

It is not clear, why the third parameter in \eq{3ps} is irrelevant.  Considering the nature of the problem, we may suggest two possible candidates for $E_s$: $E_s^{(1)}=\sigma/er_c$ and $E_s^{(2)}=e/r^2_c$.  In fact, if we use $E_s^{(2)}$ in \eq{3ps}, then $E/E_s^{(2)}$ does not depend on $E$, and this is exactly the universal behavior of \eq{2ps}, but, again, why one should use $E_s^{(2)}$ instead of $E_s^{(1)}$ is unclear.

Typically, the universality of transients is considered as an inherent property of the dispersive transport. In our case validity of \eq{2ps} means that the mean carrier velocity (and, hence, the mobility) does not depend on $L$ and, in this respect, the transport is also non-dispersive.

\begin{figure}[tbh]
     \begin{center}
\includegraphics[width=3in]{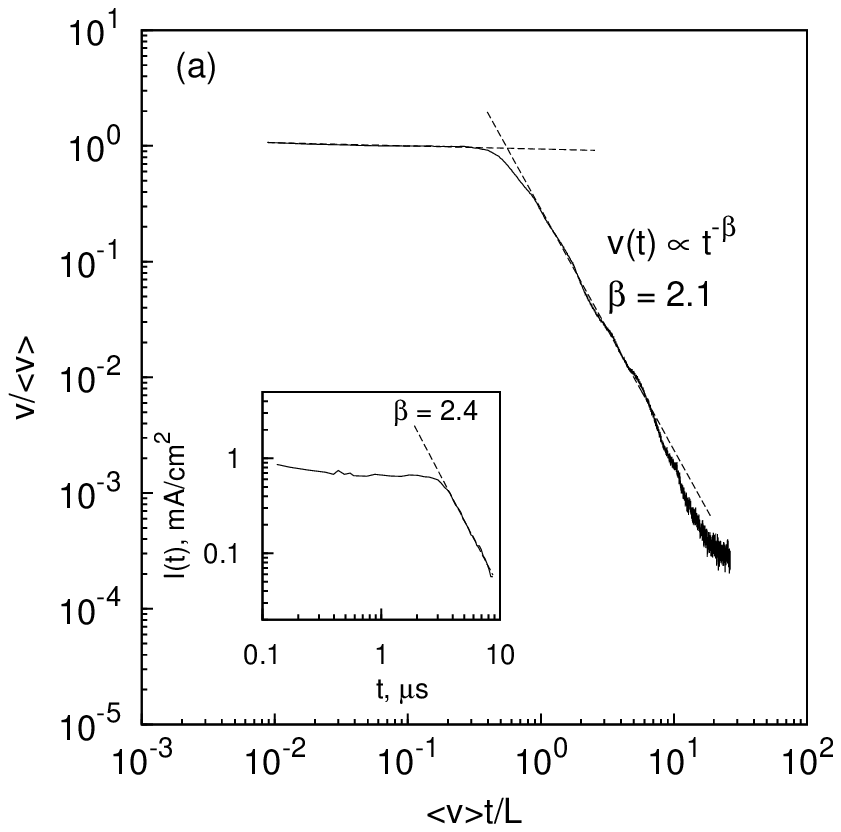}
\includegraphics[width=3in]{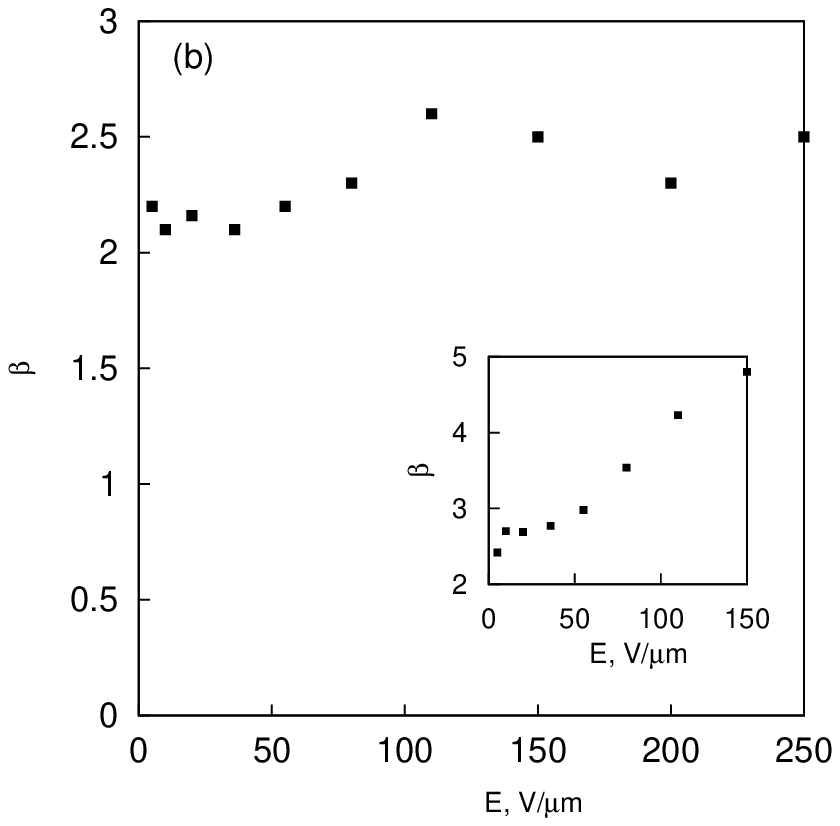}
     \end{center}
\caption{Typical shape of the transient for $E=10$ V/$\mu$m, $L=20\thinspace000\thinspace a$, inset shows the transient for the fluorene-arylamine copolymer, experimental data is borrowed from Fig. 2 in Ref. \onlinecite{Poplavskyy:415} (a), and the dependence of the parameter $\beta$ in \eq{tail} on the electric field, inset shows the corresponding dependence for 100\% filled equivalent lattice (b).\label{beta}}
\end{figure}
%===== fig5

\begin{figure}[tbh]
     \begin{center}
\includegraphics[width=3in]{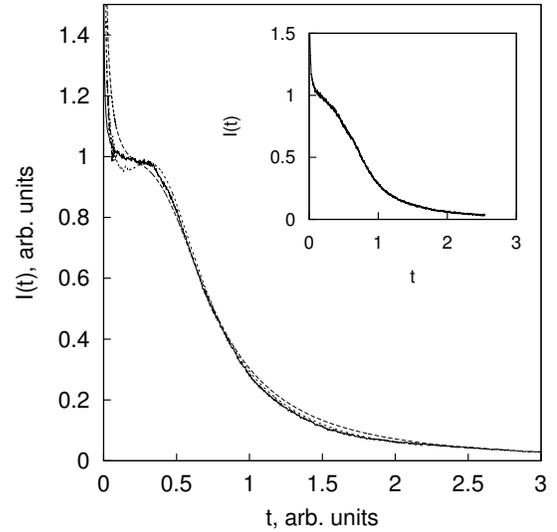}
     \end{center}
\caption{Simulated transient for $E=36$ V/$\mu$m and $L=20\thinspace000\thinspace a$ (solid line) and transients 3 and 4 from Fig. 1 in Ref. \onlinecite{Tyutnev:115107} (broken and dotted lines, correspondingly); both experimental transients have been measured at $E=40$ V/$\mu$m but for slightly different thickness of the transport layer (14$\mu$m and 15$\mu$m). Experimental transients were re-scaled to obtain the best fit. Inset shows the more dispersive GDM transient with no visible plateau for the same $E$.\label{trans}}
\end{figure}
%===== fig6

The most interesting feature is the behavior of the tail of the transients. We found that at room temperature the tail obeys the law
\begin{equation}\label{tail}
    v(t)\propto 1/t^{\beta},
\end{equation}
with parameter $\beta$ being very close to 2 (Fig. \ref{beta}). It is quite probable that $\beta$ is even more close to 2 because determination of the tail parameters is  very sensitive to inevitable statistical errors. For the equivalent 100\% filled lattice $\beta$ strongly depends on $E$ and transients are not universal (Fig. \ref{beta}b, inset).

Close values of $\beta$ are observed for other organic glasses, too (see inset in Fig. \ref{beta}a for the transient in fluorene-arylamine copolymer at room temperature and $E=1.9\times 10^5$ V/cm).\cite{Poplavskyy:415} For this particular glass the universality with respect to $E$ takes place, too, and the transients demonstrate well-defined plateau indicating the non-dispersive transport. In Ref. \onlinecite{Schein:175} it was found that $\beta\approx 3$ and again is independent of $E$ and $L$. Note, though, that the the actual value of $\beta$ is sensitive to the time range, available for the analysis. Typically, in experiments the tested time range is not very wide and the transients decay just by one order of magnitude (see inset in Fig. \ref{beta}a). In this situation true values of $\beta$ for very long times might be even more close to 2.

It is worth noting that the power law decay of transients has been observed even in liquid crystals, and again it was preceded by a well-defined plateau,\cite{Duzhko:113312} though in that case  $\beta\approx 3$ or 4, depending on the particular liquid crystalline phase. Hence, even in much more locally ordered materials the tail of the transient does not follow the exponential law, inherent for classical diffusion.

Power law decay of the transient is typically associated with dispersive transport,\cite{Scher:2455} where
\begin{equation}\label{dispersive}
I(t)\propto\begin{cases}
    t^{-(1-\alpha)}, \hskip10pt t < t_T,\\t^{-(1+\alpha)}, \hskip10pt t > t_T,
\end{cases}
\end{equation}
where $t_T$ is some characteristic transport time and $0 \le \alpha \le 1$. If $\beta\approx 2$, then $\alpha=\beta-1\approx 1$. Formally, this means that we have the boundary case between dispersive and non-dispersive transport, and the carrier mobility does not depend on the layer thickness $L$, because $\mu \propto L^{1-1/\alpha}=L^{0}=\textrm{const}$. Hence, the power law decay of the transients with $\beta\approx 2$ and independence of $\left<v\right>$ on $L$ are closely related and mutually consistent phenomena.

We checked an agreement between shapes of the simulated transients and experimental ones, measured in 30\% DEH:PC layers.\cite{Tyutnev:115107} Result of the comparison is shown in Fig. \ref{trans}. Difference between values of $E$ (and, hence, an average carrier velocity) and $L$ in experiment and simulation ($L=15.4$ $\mu$m) is not important due to transient universality: in this procedure we compare only the relative shapes of the transients.

\begin{figure}[tbh]
     \begin{center}
\includegraphics[width=3in]{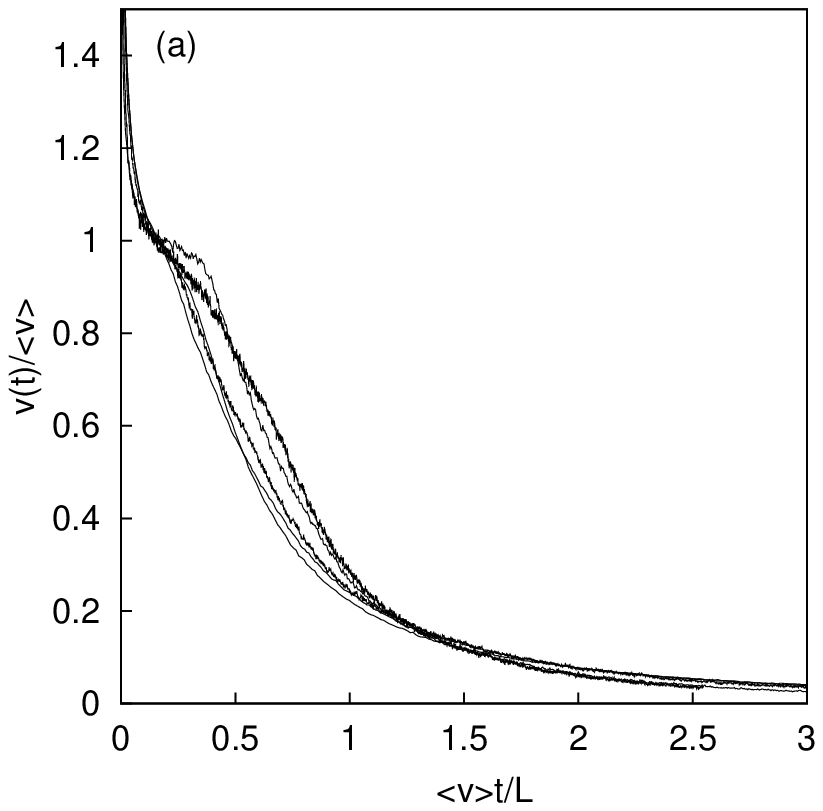}
\includegraphics[width=3in]{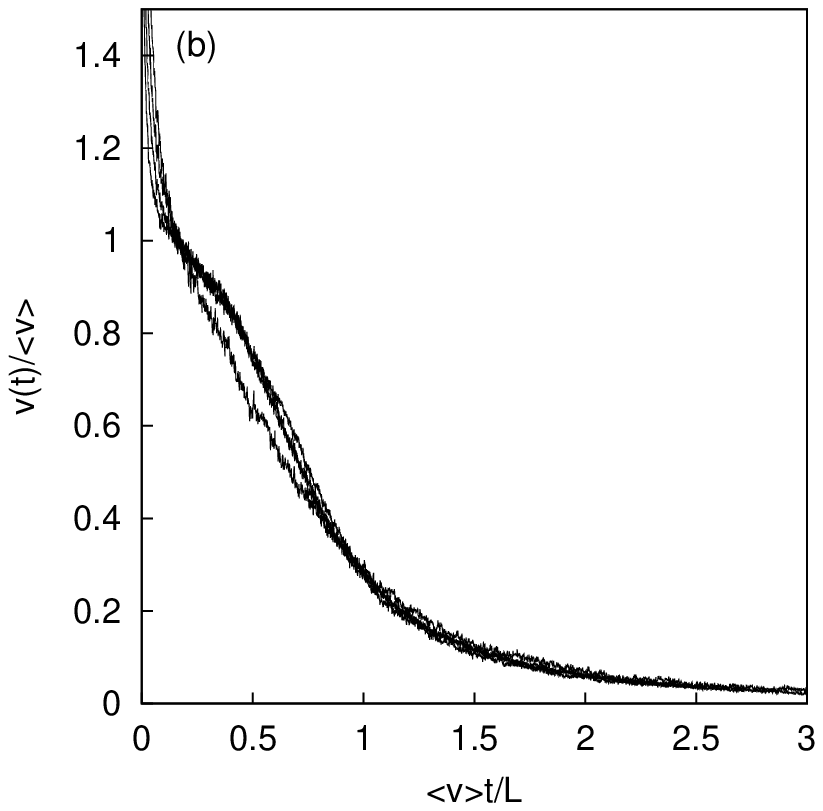}
     \end{center}
\caption{Check of the universality for the GDM transients: transients for electric field in the range $5-200$ V/$\mu$m and $L=20\thinspace000\thinspace a$ (a); transients for layer thickness in the range $2\thinspace000\thinspace a$ -- $20\thinspace000\thinspace a$ and $E=36$ V/$\mu$m (b).\label{GDMu}}
\end{figure}
%===== fig7

\begin{figure}[tbh]
     \begin{center}
\includegraphics[width=3in]{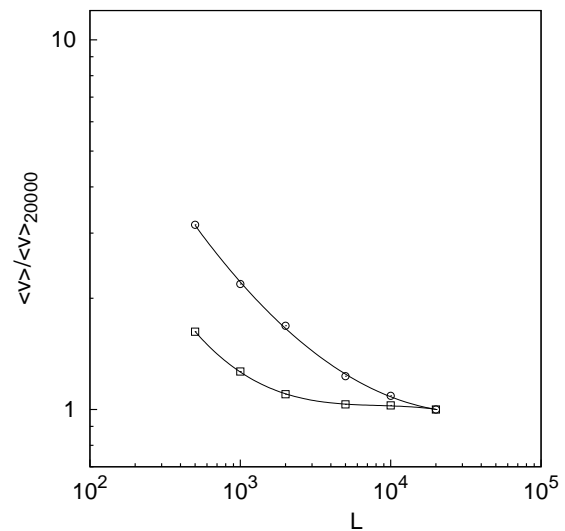}
     \end{center}
\caption{Dependence of the mean carrier velocity $\left<v\right>$ on the thickness $L$ (number of lattice planes) of the transport layer for $E=55$ V/$\mu$m in the GDM ($\circ$) and DG model ($\Box$), correspondingly. Velocity is normalized by the corresponding velocity $\left<v\right>_{20000}$ for $20\thinspace000$ lattice planes. Lines are provided as guides for an eye.\label{v(L)}}
\end{figure}
%===== fig8

The GDM shows poorer universality in the same field and thickness range (Fig. \ref{GDMu}). Actually, for most values of $E$ the GDM transients are so dispersive that reliable determination of the transit time in double linear coordinates and, hence, calculation of $W$ is not possible. This is a very typical situation: if we  extract the relevant transport parameters for the GDM from the experimental data and then use these parameters for the simulation of the GDM transients, the output of the simulation produces much more dispersive transients than the experimental ones.\cite{Tyutnev:325105} Dispersive regime for the GDM transport can be illustrated by the dependence of the mean carrier velocity $\left<v\right>$ on $L$ (see Fig. \ref{v(L)}). Mean velocity monotonously decreases with $L$ for the GDM, while for the DG model it is almost a constant for $L \ge 2\thinspace000\thinspace a$. This observation supports the conception of the breakdown of the this particular type of universality for the GDM, because for the validity of \eq{2ps} the mean velocity should be independent of $L$. At the same time, dispersive transport in the GDM and DG model is not described by \eq{dispersive}, because in both cases the corresponding values of $\alpha$ are not equal for $t < t_T$ and $t>t_T$. This fact was already mentioned by B{\"a}ssler.\cite{Bassler:15} Fig. \ref{v(L)} provides additional support to unusual dispersion of transients  in organic glasses, because if \eq{dispersive} is valid, then $\left<v(L)\right>\propto L^{1-1/\alpha}$, which does not agree with Fig. \ref{v(L)}.

\section{Conclusion}

We carried out Monte Carlo simulation of the hopping charge transport in the DG model with parameters ($\sigma$ and $\gamma a$) directly taken from the experimental data for the archetypical molecular doped polymer 30\% DEH:PC and assuming the simplest independent random distribution of DEH transport molecules at the sites of a cubic lattice. We found that the mobility field dependence for moderate field in the DG model for $c=0.3$ retains its PF form, in good agreement with the general picture of the hopping transport in the correlated energy landscape. The most important parameter, namely slope of the mobility field dependence, was found to be very close to the experimental one.

Hence, for the DG model the simulation that uses the experimental parameters, derived from the temperature dependence of the low-field mobility ($\sigma$) and dependence of low field mobility on concentration of transport sites ($\gamma a$), is capable to reproduce well the mobility field dependence and shape of the transients. This agreement becomes even more convincing if one takes into account that for the model of equivalent 100\% filled lattice the slope of the mobility field dependence and shape of transient are in striking disagreement with the experiment (see Figs. \ref{PF_exp} and \ref{beta}). And then, the simplest natural modification of the model (i.e., the use of partially filled lattice and assumption of the random independent spatial distribution of transport sites) immediately brings simulated curves very close to the experimental ones. There is no need to introduce additional spatial disorder apart from the random distribution of transport site.

We would like to emphasize an importance of this fact because recent survey suggested that the origin of the energetic disorder in molecular doped polymers and very nature of the activation energy of the hopping transport in these materials should be reconsidered.\cite{Schein:7295} According to this study, activation energy of the hopping transport in molecularly doped polymers is mostly of the intramolecular origin. At the same time, there is a general belief that the energetic disorder with the magnitude $\sigma\simeq 0.1$ eV is typical for amorphous organic materials. Moreover, long range spatial correlation of the energy landscape is, at the moment,  the only possible candidate for the explanation of the PF mobility field dependence.\cite{Dunlap:542,Novikov:2584} Our simulation shows that it is possible to reproduce a typical field dependence in polar organic material (i.e., the dipolar glass with $\sigma\approx 0.13$ eV) without using adjustable parameters and taking only a simplest assumption of the random distribution of hopping sites in the bulk of organic material. In this simulation the magnitude of the slope of the mobility dependence is directly dictated by the magnitude of the dipolar disorder. Hence, the good agreement between our simulation and the experiment \cite{Mack:7500,Schein:1067,Tyutnev:115107} provides an addition support in favor of the existence of correlated energetic disorder with $\sigma\simeq 0.1$ eV in organic materials. At the same time, as it was stressed in Ref. \onlinecite{Schein:7295}, in polymers, doped with highly polar transport dopants, the magnitude of the total disorder, which presumably includes a significant and variable dipolar contribution, and calculated using the temperature dependence of the low field mobility, seemingly does not depend on the dopant concentration. This disagreement, why the dipolar disorder clearly reveals itself in the mobility field dependence and yet its contribution to the mobility temperature dependence is hidden, poses a major puzzle for the problem of carrier transport in amorphous organic materials. Moreover, recent study casts doubt on the validity of the use of correlated models for description of the charge transport in polymers, in contrast with the transport properties of low molecular weight materials.\cite{Vries:163307} The analysis was based on rather indirect comparison of the measured current-voltage curves with the predicted ones. At the same time, much more simple and direct TOF experiments unambiguously demonstrate that mobility field dependence agrees with predictions of the correlated model in both polymers and low molecular weight glasses, without any qualitative difference.\cite{Borsenberger:9} Quite possibly, that the disagreement found in Ref. \onlinecite{Vries:163307}, also stems from the implicit use of the lattice gas model. New experiments on the concentration dependence of transport parameters with careful analysis of the transients along the lines suggested in Ref. \onlinecite{Novikov:2532}, as well as more thorough study of the simulated mobility dependence on the concentration of transport sites, are extremely desirable. At the same time, further comparison of the simulation data with experimental ones could be considered as a logical extension of our study. Possible tests may include attempts to describe transformation of transients and variation of the slope of the mobility field dependence with temperature and concentration of transport molecules. A major difficulty for the realization of that programme is the scarcity of raw transient data in the literature.\cite{Novikov:2532}

We showed also that at the room temperature transients demonstrate universality with respect to $L$ and $E$, and are non-dispersive according to the usual definition (carrier mobility does not depend on $L$, and the transient demonstrates a flat plateau if plotted in double linear $I$ vs $t$ coordinates). Yet the shape of the transients differs significantly from the result of the simple diffusive approximation, which was suggested for the description of experimental TOF transients.\cite{Hirao:1787,Hirao:4755,Nishizawa:L250} Two major differences are the universality of the transients with respect to $L$ and power law decay of the tail. It is worth  noting that universality with  respect to $L$ naturally arises in the model where charge transport is described by the broad distribution of the effective carrier velocities,\cite{Novikov:68,Scott:8603} but universality with respect to $E$ cannot be explained by this approach. It is also worth to note that the development of a flat plateau and simultaneous power law tail of the current with $\beta\simeq 2$ cannot be described by the multiple trapping model, frequently used for the description of the hopping transport in random media.\cite{Hartenstein:8574}

\section*{Acknowledgements}
SVN is grateful for partial financial support from the RFBR grants 10-03-92005-NNS-a and 11-03-00260-a, and  from the Russian Ministry of Education and Science (state contract 16.523.11.3004).

\end{document}